\documentclass[journal=jacsat,manuscript=article]{achemso}

\usepackage{amsmath,amssymb} % for math symbols
\usepackage[normalem]{ulem}
\usepackage{braket}
\usepackage{dsfont}
\usepackage{graphicx}
\usepackage{dcolumn}
\usepackage{bm}
\usepackage{color}
\usepackage{upgreek}
\usepackage{multirow}

\usepackage[dvipsnames]{xcolor}

\title{Quenching of excitons at grain boundaries in C$_{60}$ thin films}

\author{Rysa Greenwood}
\affiliation{Department of Physics and Astronomy, University of British Columbia, Vancouver, British Columbia, V6T 1Z1 Canada}
\alsoaffiliation{Quantum Matter Institute, University of British Columbia, Vancouver, British Columbia, V6T 1Z4 Canada}

\author{Bradley G. Guislain}
\affiliation{Department of Physics and Astronomy, University of British Columbia, Vancouver, British Columbia, V6T 1Z1 Canada}
\alsoaffiliation{Quantum Matter Institute, University of British Columbia, Vancouver, British Columbia, V6T 1Z4 Canada}

\author{MengXing Na}
\affiliation{Department of Physics and Astronomy, University of British Columbia, Vancouver, British Columbia, V6T 1Z1 Canada}
\alsoaffiliation{Quantum Matter Institute, University of British Columbia, Vancouver, British Columbia, V6T 1Z4 Canada}

\author{Alexandra B. Tully}
\affiliation{Department of Physics and Astronomy, University of British Columbia, Vancouver, British Columbia, V6T 1Z1 Canada}
\alsoaffiliation{Quantum Matter Institute, University of British Columbia, Vancouver, British Columbia, V6T 1Z4 Canada}

\author{Sergey Zhdanovich}
\affiliation{Quantum Matter Institute, University of British Columbia, Vancouver, British Columbia, V6T 1Z4 Canada}

\author{Jerry Icban Dadap}
\affiliation{Quantum Matter Institute, University of British Columbia, Vancouver, British Columbia, V6T 1Z4 Canada}

\author{Sydney K. Y. Dufresne}
\affiliation{Department of Physics and Astronomy, University of British Columbia, Vancouver, British Columbia, V6T 1Z1 Canada}
\alsoaffiliation{Quantum Matter Institute, University of British Columbia, Vancouver, British Columbia, V6T 1Z4 Canada}

\author{Vanessa King}
\affiliation{Department of Chemistry, University of British Columbia, Vancouver, British Columbia, V6T 1Z1 Canada}
\alsoaffiliation{Quantum Matter Institute, University of British Columbia, Vancouver, British Columbia, V6T 1Z4 Canada}

\author{Jiabin Yu}
\affiliation{Department of Physics and Astronomy, University of British Columbia, Vancouver, British Columbia, V6T 1Z1 Canada}
\alsoaffiliation{Quantum Matter Institute, University of British Columbia, Vancouver, British Columbia, V6T 1Z4 Canada}

\author{Giorgio Levy}
\affiliation{Quantum Matter Institute, University of British Columbia, Vancouver, British Columbia, V6T 1Z4 Canada}

\author{Arthur K. Mills}
\affiliation{Quantum Matter Institute, University of British Columbia, Vancouver, British Columbia, V6T 1Z4 Canada}

\author{Matteo Michiardi}
\affiliation{Quantum Matter Institute, University of British Columbia, Vancouver, British Columbia, V6T 1Z4 Canada}

\author{Andrea Damascelli}
\affiliation{Department of Physics and Astronomy, University of British Columbia, Vancouver, British Columbia, V6T 1Z1 Canada}
\alsoaffiliation{Quantum Matter Institute, University of British Columbia, Vancouver, British Columbia, V6T 1Z4 Canada}

\author{Sarah A. Burke}
\email{saburke@phas.ubc.ca}
\affiliation{Department of Physics and Astronomy, University of British Columbia, Vancouver, British Columbia, V6T 1Z1 Canada}
\alsoaffiliation{Quantum Matter Institute, University of British Columbia, Vancouver, British Columbia, V6T 1Z4 Canada}
\alsoaffiliation{Department of Chemistry, University of British Columbia, Vancouver, British Columbia, V6T 1Z1 Canada}

\author{David J. Jones}
\email{djjones@phas.ubc.ca}
\affiliation{Department of Physics and Astronomy, University of British Columbia, Vancouver, British Columbia, V6T 1Z1 Canada}
\alsoaffiliation{Quantum Matter Institute, University of British Columbia, Vancouver, British Columbia, V6T 1Z4 Canada}

\begin{document}

\begin{abstract}
Exciton lifetimes play a critical role in the performance of organic optoelectronic devices. In this work, we investigate how the presence of multiple rotational domains, and therefore grain boundaries, impacts exciton dynamics in thin films of C$_{60}$/Au(111) using time and angle-resolved photoemission spectroscopy (TR-ARPES). We find that films with multiple rotational domains exhibit shorter exciton lifetimes and evidence of exciton-exciton annihilation, even when one domain predominates. Scanning tunneling microscopy (STM) measurements reveal electronic structure changes resulting from a locally reduced dielectric constant at grain boundaries, providing a mechanism for lifetime reduction through exciton funneling and other additional decay channels. These findings highlight the critical role of film quality in determining intrinsic exciton lifetimes, and show that minuscule amounts of disorder that are nearly undetectable by ensemble measurements can significantly impact dynamics. These results imply that precise structural control is essential for optimize the performance of organic optoelectronic devices.

\end{abstract}

\maketitle

%\section{Introduction}

The efficiency of organic photovoltaic (OPV) devices is directly linked to the ability of excitons to dissociate into free charges within organic semiconductors \cite{mpelzer_charge_2016}. As exciton dissociation typically occurs at donor–acceptor interfaces, the likelihood of successful charge separation depends strongly on whether the exciton can reach a suitable interface before undergoing recombination\cite{mpelzer_charge_2016}. Accordingly, two key parameters that impact device performance are the exciton diffusion length and lifetime \cite{vmikhnenko_exciton_2015, mpelzer_charge_2016}. Previous studies have demonstrated that structural disorder, such as grain boundaries or amorphous domains, can significantly reduce exciton diffusion lengths in organic thin films\cite{terao_correlation_2007, fravventura_determination_2012, vmikhnenko_exciton_2015}. Additionally, photoluminescence studies show the exciton peak is quenched in some organic semiconductors with increasing structural disorder\cite{lunt_relationship_2010, sajjad_controlling_2015}, indicating that an increase in disorder opens up non-radiative decay pathways for the exciton. In C$_{60}$, structural defects known as X-traps have been linked shifts in energy of the fluorescence spectra and local relaxation of selection rules\cite{guss_fluorescence_1994,merino_exciton_2015, roslawska_single_2018, grose_nanoscale_2016}, resulting in localization of excitons in 0-dimensions with potential applications as single photon emitters\cite{roslawska_single_2018, grose_nanoscale_2016}. 

In this work, we investigate the role of 1-dimensional grain boundaries on charge transfer exciton lifetimes, by studying three epitaxial C$_{60}$ films on a Au(111) single crystal that exhibit different proportions of dominant and secondary rotational domains, and consequently different proportions of grain boundaries between these domains. Low-energy electron diffraction (LEED) reveals subtle differences in these films that are only observed in strongly saturated LEED images, indicating the presence of a strongly dominant domain with small proportions of other, less favored structures \cite{tully_two-stage_2024, veenstra_interface_2002, tjeng_development_1997}. Next, we compare the lifetime of an excited state for three different samples using time- and angle-resolved photoemission spectroscopy (TR-ARPES). Our results show that even a small proportion of multi-domain structure significantly reduces lifetimes of the excited states. To better understand the behavior at the grain boundaries, we perform scanning tunneling microscopy (STM) and spectroscopy (STS) measurements. Based on our observations we infer the two-particle states are lower in energy at grain boundaries and we attribute this shift to the disruption of the close-packed structure of C$_{60}$. We conclude the grain boundaries act as exciton funnels, attracting nearby excitons and increasing the local exciton density. The faster dynamics observed in films with grain boundaries are thus attributed to both: (i) increased  radiative and non-radiative decay due to local symmetry lowering;
%increased phonon-assisted decay pathways resulting from broadened phonon spectrum; 
and (ii) increased exciton-exciton annihilation at a grain boundary. While the effect of grain boundaries on exciton lifetimes has been studied extensively in transition-metal dichalcogenides\cite{man_functional_2021, kim_biexciton_2016, kim_recent_2023, mueller_exciton_2018, cunningham_charge_2016, zhou_grain_2019}, their effects remain comparatively underexplored in C$_{60}$ and organic semiconductors in general. These findings not only highlight channels for performance loss in OPV devices but also the ability of C$_{60}$ to act as an exciton funnels, with potential applications as single-photon emitters \cite{tonndorf_single-photon_2015}.

%\section{Results and Discussion}

\begin{figure}
	\centering
	\includegraphics[scale=0.9]{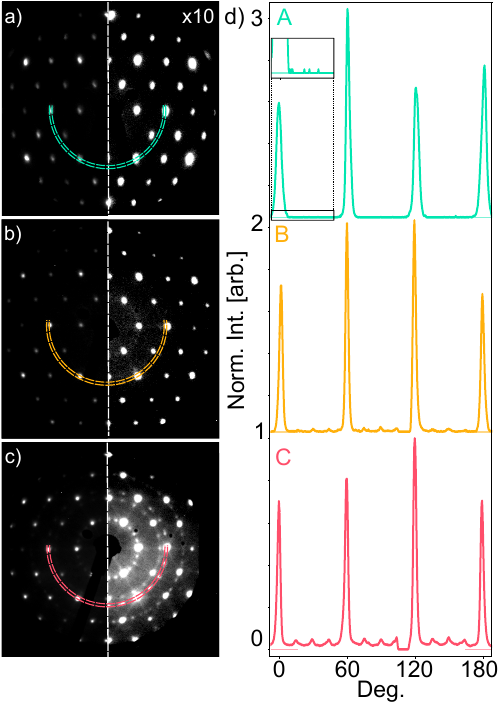}
	\caption{LEED images (a)-(c) comparing three different 10~ML samples of C$_{60}$/Au(111). The intensity is shown unsaturated on the left, and is saturated by a factor of 10 on the right side of each image to highlight the presence (or lack of) additional rotational domains. (d) Line profiles are taken along a semi-circular path, marked by dotted lines through: (a) Sample A, green; (b) Sample B, orange; and (c) Sample C, red. Four major diffraction peaks in the line profiles are signatures of the dominant 2$\sqrt{3}\times$2$\sqrt{3}$ R30~$^\circ$ rotational domain of C$_{60}$/Au(111).  Three minor peaks are present between the major diffraction peaks for Samples B and C indicating presence of at least three additional rotational domains. The expanded inset for Sample A shows no structure corresponding to additional domains in this same region. Each line profile is normalized between 0 and 1. Line profile from Sample B and C are shifted by 1.1 and 2.1, respectively. The noise floor of the LEED measurement is marked for each curve using a semitransparent line on the left and right side of each curve in panel (d).}
	\label{fig:LEED}
\end{figure}

%\subsection{Static Characterization using LEED and ARPES}
LEED is used to establish the crystallinity of films grown by molecular beam epitaxy (MBE).  Additional rotational domains in the sample manifest as multiple single-domain LEED patterns rotated relative to one another. Figure~\ref{fig:LEED} compares LEED images taken with a beam energy of 15~eV for three different 10 monolayer (ML) samples of C$_{60}$/Au(111), using a spot size of $\sim250\mu m$. In the unsaturated images on the left of Fig.\,\ref{fig:LEED}(a)-(c) only a single rotational domain is evident for all three samples.
%Each image is shown using an unsaturated colour scale on the left and a colour scale which is saturated by 10 times on the right to highlight the presence (or lack of) additional rotational domains. Line profiles are taken along the dotted semi-circular path shown in Fig.\,\ref{fig:LEED} a)-c) and are plotted in Fig \ref{fig:LEED}d), with Samples B and C offset in intensity by 1.1 and 2.1, respectively. Also shown are the noise floor of the LEED measurement for these curves with a semi-transparent line on the left and right side of each curve in panel d).
However, when the contrast is enhanced by 10x on the right of Fig.\,\ref{fig:LEED}(a)-(c), Samples B and C reveal at least three additional rotational domains. In contrast, Sample A exhibits only diffraction peaks consistent with a single rotational domain of C$_{60}$ previously confirmed to be the 2$\sqrt{3}\times$2$\sqrt{3}$ R30~$^\circ$ superstructure \cite{tully_two-stage_2024}. We extract line profiles from these images along the indicated semi-circular paths as the dotted lines. These profiles, shown in Fig.\,\ref{fig:LEED}(d), further support this observation and display major peaks at 0$^\circ$, 60$^\circ$, 120$^\circ$ and 180$^\circ$, which are attributed to the  2$\sqrt{3}\times$2$\sqrt{3}$ R30~$^\circ$ superstructure \cite{tully_two-stage_2024}. 
Here, the three minor peaks that appear between each pair of major peaks in the orange (Sample B) and red (Sample C) curves have intensity less then 2$\%$ and 8$\%$, respectively, confirming that the 2$\sqrt{3}\times$2$\sqrt{3}$ R30~$^\circ$ superstructure remains the predominant domain. 

Static ARPES is performed using a 21.2~eV He I line from a helium discharge lamp with a spot size of 1~mm for each sample to further investigate and compare the film quality. The resulting HOMO spectra of C$_{60}$ for each sample exhibit comparable features, highlighting the difficulty in distinguishing differences based on ARPES alone. These spectra are included in Fig.\,S2 of the supporting information. Additionally, ultraviolet photoemission spectra, shown in Fig.\,S3 of the supporting information, confirm that there is no chemical difference between samples, indicating that the only difference between these films is the presence of additional rotational domains and the accompanying grain boundaries.  

\begin{figure*}
	\centering
	\includegraphics{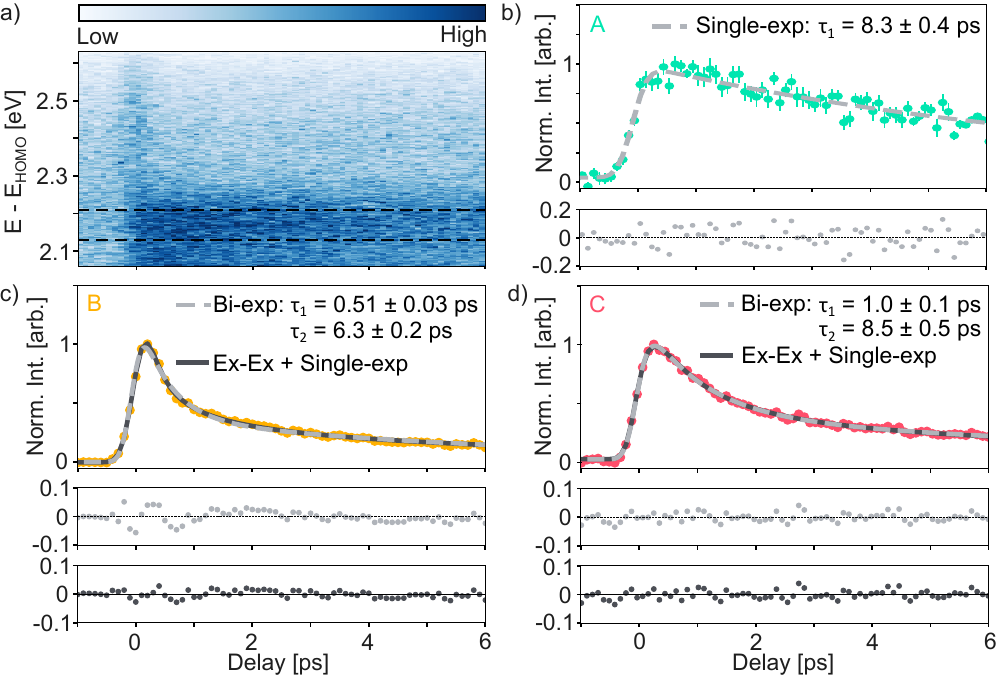}
	\caption{ Dynamics of the CT1-like excitonic state in 10~ML C$_{60}$/Au(111). (a) Angle-integrated time scan at the $\Gamma$-point for Sample A. (b)-(d) Dynamics of the CT1-like excitonic state of Sample A, B, and C respectively. The energy integration range for these curves is outlined using dotted lines in panel (a). (b) is fit using a single-exponential decay model (grey dotted line) (c) and (d) are fit using a bi-exponential decay model (grey dotted line) and a model which incorporates an exciton-exciton annihilation term (dark grey solid line). The absolute residuals for each fit are shown in the panels below the corresponding data. Measurements are at 10~K. Pump fluence is 350~$\upmu$J/cm$^2$, 300~$\upmu$J/cm$^2$ and 212~$\upmu$J/cm$^2$ for (a), (b) and (c) respectively. The time resolution for (a) is 470~fs, and (b) and (c) is 230~fs. While we plot data to 6 ps for better visualization here, the entire datasets with longer delays used in the fits appear in supporting information.}
    %Dynamics of the CT1-like excitonic state in 10~ML C$_{60}$/Au(111) at the $\Gamma$-point with $E-E_{HOMO}$ = 2.17$\pm$0.04~eV for Samples A, B and C shown in panels (a), (b), and (c) respectively.  The data for all three samples are fit with a bi-exponential decay model (light grey, dotted). In addition, Samples B and C are also fit using a model that includes an exciton-exciton annihilation term (dark grey, solid). The absolute residuals for each fit are shown in the panels below the corresponding data. Measurements are at 10~K. Pump fluence is 350~$\upmu$J/cm$^2$, 300~$\upmu$J/cm$^2$ and 212~$\upmu$J/cm$^2$ for (a), (b) and (c) respectively. The time resolution for (a) is 470~fs, and (b) and (c) is 230~fs. While we plot data to 6 ps for better visualization here, the entire datasets with longer delays used in the fits appear in supporting information.}
	\label{fig:dynamics}
\end{figure*}

%\subsection{Dynamic Characterization using TR-ARPES}

%Despite the minimal differences in LEED crystallinity measurements of these three films, the temporal dynamics of the excitons probed by TR-ARPES are significantly modified. 
Having established the structural differences between samples, we now focus on the excited states dynamics of C$_{60}$. Figure \ref{fig:dynamics}(a) depicts a time scan of Sample A, integrated about the $\Gamma$-point at an energy range which lies within the band gap of C$_{60}$. At a delay of 0~ps a broad, short lived excitation is observed around 2.4~eV, with respect to the C$_{60}$ HOMO, which decays into another broad, longer lived excitation at an energy of 2.17~eV. Energy integrated dynamics of these two states are found in the supporting information to better highlight the decay of one state into the other. This is consistent with previous observations and were both previously attributed to singlet, charge transfer exciton states referred to as CT2 and CT1 for the higher and lower lying states respectively\cite{emmerich_ultrafast_2020, bennecke_disentangling_2024, stadtmuller_strong_2019}.
%the lower lying excitation has previously been attributed to a singlet charge transfer exciton referred to as CT1\cite{emmerich_ultrafast_2020, bennecke_disentangling_2024, stadtmuller_strong_2019}.
%We observe broad exciton-like signatures at an energy of E$_{CT1}$ = 2.17~eV with respect to the C$_{60}$ HOMO, which has previously been attributed to a singlet charge transfer exciton referred to as CT1 \cite{emmerich_ultrafast_2020, bennecke_disentangling_2024, stadtmuller_strong_2019}. 
For consistency with earlier work, we will refer the population around 2.17~eV as CT1 throughout. 
%The CT1 transition is forbidden by dipole selection rules, making it highly unlikely that CT1 is being directly populated by the pump. Instead, CT1 is populated via decay from higher energy excitations on timescales shorter than the time resolution of our system (see supporting Fig.\,S8). 

Figure~\ref{fig:dynamics}(b)-(d) show the time-resolved dynamics at the $\Gamma$ point for the CT1 corresponding to Samples A, B, and C, respectively. The energy integration region for the CT1 is shown with dotted lines in Fig.\,\ref{fig:dynamics}(a), and traces are normalized between 0 and 1 to the peak intensity. To analyze the relaxation dynamics, we fit the data using a rising exponential (to incorporate the decay from the higher lying direct transition state into the CT1) and a single exponential decay. This entire model is convolved with a Gaussian to account for the instrument response. The single-exponential fit to the data from sample A is shown as a dotted grey line in panel (b) of Fig.\,\ref{fig:dynamics}, along with the absolute residuals below. We extract a lifetime of 8.3 $\pm$ 0.4~ps and the residuals are well distributed with a reduced $\chi^2_{red}$ = 1.4, indicating the dynamics can be described using a single decay component. Contrarily, the single exponential decay model does not accurately describe the data from Samples B and C, resulting in structured residuals and $\chi^2_{red}$ of 21.6 and 11.2, respectively (fit results included in supporting information Fig.\,S6). Instead, the CT1 of Samples B and C must be fit using a model with an additional decay component.
%To analyze the relaxation dynamics, we start by fitting the data using a rising exponential (to incorporate the decay from the higher lying direct transition state into the CT1) and bi-exponential decay model. This entire model is convolved with a Gaussian to account for the instrument response. The bi-exponential fits are shown as dotted grey lines in each panel of Fig.\,\ref{fig:dynamics}, along with the absolute residuals below each respective dataset. 

Towards this end, a bi-exponential decay model (including  a rising exponential as before) is convolved with a Gaussian, is used to extract a fast and slow decay component of the CT1 for Samples B and C. The fit is shown as the dotted grey line in panels (c) and (d) of Fig.\,\ref{fig:dynamics}. For Sample B we extract lifetimes of $\tau_1$ = 0.51 $\pm$ 0.03~ps and $\tau_2$ = 6.3 $\pm$ 0.2~ps, while Sample C yields slightly slower dynamics of $\tau_1$ = 1.0 $\pm$ 0.1~ps and $\tau_2$ = 8.5 $\pm$ 0.5~ps. The fit to Sample C shows well distributed residuals and reduced $\chi^2_{red}$ of 1.0. However, the best fit for Sample B using the bi-exponential model, still results in structured residuals at delays close to 0~ps and a $\chi^2_{red}$ = 2.6; an improvement from the single-exponential model but still an overall poor fit. Despite the limitations of the fit for Sample B, the extracted lifetimes clearly indicate that an additional, faster component must be considered to describe the dynamics of the CT1 in samples which contain multiple rotational domains.

Given the possible role of grain boundaries in co-localizing excitons, we develop a model which incorporates an exciton-exciton annihilation term. This model assumes two populations: one away from grain boundaries which decays linearly according to the rate equation ${dN}/{dt} = -N$, where $k$ is the linear decay rate, and another near grain boundaries which decays via exciton-exciton annihilation according to the rate equation ${dN}/{dt} = -k_{ExEx}N^2$, where k$_{ExEx}$ is the rate of exciton-exciton annihilation. Solving these rate equations individually and then summing their result as well as including  a rising term gives, 

%$\frac{dN}{dt} = -\frac{1}{\tau'}N$
%$\frac{dN}{dt} = -k_{ExEx}N^2$
\begin{equation}
N(t) = (1-e^{\frac{-t}{\tau_{rise}}})\left(A_1e^{-kt} + \frac{A_2}{1+k_{ExEx}A_2t}\right).
\end{equation}

\noindent Equation 1 is convolved with a Gaussian as before, and the resulting fits are shown in Fig.\,\ref{fig:dynamics}(c)-(d) as dark grey, solid lines. For Sample B, $k$=0.127 $\pm$ 0.006~ps$^{-1}$ and $k_{ExEx}$=3.3 $\pm$ 0.2~ps$^{-1}$. The residuals are more evenly distributed than when fit using the bi-exponential decay model and we find an improved $\chi^2_{red}$ = 1.41, indicating a better fit. For Sample C $k$=0.07 $\pm$ 0.02~ps$^{-1}$ and $k_{ExEx}$=1.3 $\pm$ 0.1~ps$^{-1}$. The residuals are similar to fitting with the bi-exponential model and we find a $\chi^2_{red}$ = 0.96. Table \ref{table:fitresults} summarizes the best fit parameters for both exponential models and the exciton-exciton annihilation plus exponential decay models. The CT1 dynamics observed in Sample B and C were also fit using a  simpler model which only allowed for an exciton-exciton annihilation term but this model could not accurately describe the dynamics (see Fig.\,S6 of the supporting information), indicating the consideration of two independent populations is necessary in samples which contain grain boundaries. Analysis of additional 10~ML C$_{60}$/Au(111) samples are also shown in the supporting information.

%To further support our claim that the dynamics of the CT1 are dependent on the number of grain boundaries, two additional 10~ML C$_{60}$/Au(111) samples are investigated and their results are consistent with Samples A, B and C (see supporting information).  Also included in the supporting information are: (i) detailed description of the fitting procedures along with fit results for all models; and (ii)  plots displaying the longer time scales (to 100~ps for Sample A and 10~ps Sample B and C).  

%A more detailed description of the fitting procedures along with all fit results is provided in the supporting information. Although the main text figures display data and fits up to 6~ps, longer time dynamics are measured and fit up to 100~ps (Sample A) and 10~ps (Sample B and C) and are also included in the supporting information. To further support our claim that the dynamics of the CT1 are dependent on the number of grain boundaries, two additional 10~ML C$_{60}$/Au(111) samples are investigated and their results are consistent with Samples A, B and C (see supporting information).  

\begin{table*}
\caption{Summary of fit results. Sample A is fit using a single-exponential decay model. Samples B and C contain two decay components and are fit using both a bi-exponential decay model and a model that incorporates an exciton-exciton annihilation term plus a single-exponential decay term.}
\begin{center}
\begin{tabular}{ |c|c|c|c|c| }
\hline
 \multicolumn{2}{|c|}{}  & {Sample A} & {Sample B} & {Sample C} \\ 
 \hline 
  \multirow{5}{*}{\shortstack{Exponential \\ decay}} &\bf{$\chi^2_{Red}$} & {1.4}&{2.6} &{1.0} \\ 
  \cline{2-5}
 &{$\tau_1$ [ps]} & {8.3 $\pm$ 0.4}&{0.51 $\pm$ 0.03} &{1.0 $\pm$ 0.1} \\
 \cline{2-5}
&{$k_1$ [ps$^{-1}$]} & { 0.121 $\pm$ 0.005 }&{1.9$\pm$ 0.1} &{1.0 $\pm$0.1} \\ 
 \cline{2-5}
&{$\tau_2$} [ps]& {\textemdash}&{6.3 $\pm$ 0.2} &{8.5 $\pm$ 0.5} \\ 
 \cline{2-5}
&{$k_2$  [ps$^{-1}$]} & {\textemdash}&{0.155 $\pm$ 0.005} &{0.114$\pm$ 0.007} \\ 

 \hline 
\hline
  \multirow{4}{*}{\shortstack{Exciton-Exciton +\\exponential decay}} &\bf{$\chi^2_{Red}$} & \bf{\textemdash}&{1.41} &{0.96} \\ 
  \cline{2-5}
 &{$\tau'$ [ps]} & \bf{\textemdash}&{7.9 $\pm$ 0.4} &{15 $\pm$ 3} \\
 \cline{2-5}
&{$k$  [ps$^{-1}$]} & \bf{\textemdash}&{0.127 $\pm$ 0.006} &{0.07 $\pm$ 0.02} \\ 
 \cline{2-5}
&{$k_{ExEx}$  [ps$^{-1}$]} & \bf{\textemdash}&{3.3 $\pm$ 0.2} &{1.3 $\pm$ 0.1} \\ 

  \hline 
\end{tabular}
\end{center}
\label{table:fitresults}
\end{table*}

With evidence from the fits for two-exciton processes, we can now rationalize the different lifetimes for Samples B and C by considering pump fluence in addition to grain boundaries. For technical reasons, the pump fluence varies between measurements and it is 350~$\upmu$J/cm$^2$, 300~$\upmu$J/cm$^2$ and 212~$\upmu$J/cm$^2$ for Sample A, B and C respectively. So while LEED data for Sample C indicates the largest proportion of additional domains in Fig\,\ref{fig:LEED}, and presumed grain boundaries, the faster dynamics of Sample B may be explained by the higher pump fluence. Moreover, fluence dependent measurements of the CT1 of Sample B, included in supporting information Fig.S9, indicate sensitivity to two-exciton processes at a fluence of 300~$\upmu$J/cm$^2$. In addition, Sample A measurements are preformed using the highest fluence, yet show no sign of two-exciton processes, indicating that the presence of grain boundaries in the C$_{60}$ films strongly increases susceptibility to two-exciton decay processes like exciton-exciton annihilation.

\begin{figure*}
	\centering
	\includegraphics{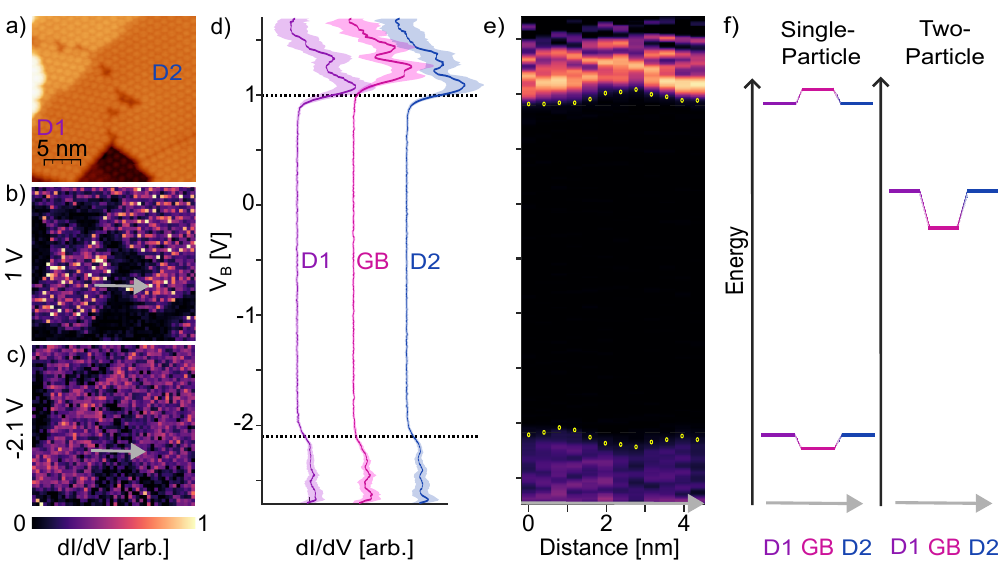}
	\caption{STM results displaying changes in the energy landscape across a grain boundary between two rotational domains of $C_{60}$.  (a) Topographic image ($V_B$=\textminus2.5~V, $I$=2~pA) in the region of a typical grain boundary in C$_{60}$/Au(111), showing mostly bi-layer  coverage with an underlying Au step, a patch of ML C$_{60}$/Au(111) at the bottom edge, and a small amount of tri-layer at the left edge. (b) and (c) dI/dV maps of same region shown in panel (a) at approximate onset of the LUMO (1~V) and HOMO (\textminus2.1~V), respectively. Scale is the same as panel (a). (d) Averaged dI/dV curves from sampled regions in D1 and D2 labelled in panel (a) and along the grain boundary. Dotted lines mark the approximate HOMO and LUMO onsets highlighting the widening of the gap along the grain boundary. (e) dI/dV colormap with same scale as panel (b) and (c) along a line profile taken across the grain boundary, with the path taken indicated by a grey arrow in panels (b) and (c). (f) A qualitative sketch of the measured single particle states and inferred changes in two-particle (excitonic) states following a path across the grain boundary similar to the grey arrow.}
	\label{fig:STM}
\end{figure*}

%\subsection{Grain Boundary Characterization using STM}
To gain further perspective of the impacts grain boundaries impose on the excitons, we use STM to investigate how the electronic structure changes across a grain boundary between two different well-ordered rotational domains. Figure~\ref{fig:STM}(a) shows an example of a grain boundary between the 2$\sqrt{3}\times$2$\sqrt{3}$ R30~$^\circ$ (denoted D1) and the 7$\times$7 R14~$^\circ$ (denoted D2) superstructures of C$_{60}$/Au(111). This image shows predominantly bilayer C$_{60}$ with a step in the underlying Au(111), a patch of ML C$_{60}$/Au(111) at the bottom edge, and a small amount of trilayer C$_{60}$/Au(111) near the left edge. The close-packed structure of C$_{60}$ is disrupted at the grain boundary to accommodate the change in rotational domain. 

Scanning tunneling spectroscopy (STS) and differential conductance maps for the bias voltage, $V_B$, near the expected LUMO (1~V) and HOMO (\textminus2.1~V) onsets are shown in Fig.\,\ref{fig:STM}(b) and (c), respectively. These maps reveal an increased gap in the dI/dV signal along the grain boundary and at step edges, and can be understood by considering the local polarizability of the environment\cite{cochrane_pronounced_2015} that is modified at the grain boundary. The transport gap is then reduced by the polarization energy  P\cite{sharifzadeh_relating_2015}, 

\begin{equation}
E_{BG} = E_G^{gas} - 2P
\end{equation}

\noindent where $E_G^{gas}$ is the HOMO-LUMO gap of C$_{60}$ in the gas phase and

\begin{equation}
P = \frac{e^2}{2a}\left( 1-\frac{1}{\epsilon}\right)  
\end{equation}

\noindent is the classical expression for the polarization energy of a point charge inside a dielectric \cite{bounds_polarization_1979,tsiper_charge_2001, tsiper_electronic_2003} where, $e$ is the elementary charge, $\epsilon$ is the effective dielectric constant, and $a$ is the radial distance from the charge to the dielectric. A change in the local dielectric environment therefore leads to a change in $E_{BG}$,

\begin{equation}
\Delta E_{BG} = \frac{e^2}{2a}\left(\frac{1}{\epsilon_2}-\frac{1}{\epsilon_1}\right)
\end{equation}

\noindent A decrease in the local dielectric constant is therefore expected to widen $E_{BG}$. We confirm this widening along the grain boundary by sampling and averaging dI/dV curves from D1, D2, and along the grain boundary (GB) which is shown in Fig.\,\ref{fig:STM}(d). As expected, the positions of the HOMO and LUMO shift to lower and higher bias voltages in STS, respectively. For additional information about the regions and the distribution of the dI/dV curves in the sampled regions see supporting information Fig.\,S10.

Further insight can be gained by plotting the dI/dV as an intensity plot along a line profile crossing the grain boundary. The line profile is shown as grey arrow in Fig.\,\ref{fig:STM}(b) and (c) with dI/dV displayed in Fig.\,\ref{fig:STM}(e). The dI/dV curves have been scaled relative to one another to improve visualization. Yellow circles mark the $V_B$ when the dI/dV signal reaches 10$\%$ of the peak signal, highlighting the smooth shift of the HOMO and LUMO to lower and higher bias voltage, respectively, as we move across the grain boundary. Several grain boundaries between different rotational domains were also studied and all showed similar results. 

Although STS measures the single-particle states (electron and hole excitations),  we can infer how the two-particle (excitonic) states are affected at the grain boundary from these measurements. The binding energy of the exciton, $E_B$, also depends on the dielectric constant,

\begin{equation}
E_B = \frac{\mu e^4}{2(4\pi\epsilon_0\epsilon\hbar)^2}
\end{equation}

\noindent where $\epsilon_0$ is the vacuum permittivity and $\mu$ is the reduced effective mass of the exciton. The change in $E_B$ due to a change in the dielectric constant is given by,

\begin{equation}
\Delta E_B = \frac{\mu e^4}{2(4\pi\epsilon_0\hbar)^2}\left(\frac{1}{\epsilon_2^2}-\frac{1}{\epsilon_1^2}\right).
\end{equation}

\noindent For realistic values of $a$ and $\mu$, $\Delta E_B$ increases more rapidly than $\Delta E_{BG}$. As a result, exciton states at the grain boundary are expected to be lower in energy due to less efficient screening, which enhances the exciton binding energy $E_B$. A qualitative sketch of the single-particle states and inferred two-particle states is illustrated in Fig.\,\ref{fig:STM}(f). 

This energetic landscape implies that the grain boundary can act as an exciton funnel, drawing in nearby excitons as it is energetically more favorable for them to reside at the grain boundary. Localization along the grain boundary results in an increase of the local exciton population and therefore  increase the probability of two-exciton processes, including exciton-exciton annihilation. These findings resemble previous work on X-traps in C$_{60}$ which show downward energy shifts of localized fluorescence spectra \cite{guss_fluorescence_1994,roslawska_single_2018, grose_nanoscale_2016}. Furthermore, the disruption of crystal symmetry at the grain boundary may increase the likelihood of additional radiative and non-radiative decay channels by relaxing the local selection rules \cite{grose_nanoscale_2016, roslawska_single_2018} and locally broadening the phonon spectrum respectively. Together, these factors are consistent with the faster dynamics and increased susceptibility to exciton-exciton annihilation in samples exhibiting multiple rotational domains and further support our fitting models employed previously. 

%\section{Conclusion}
We have shown that lifetimes of exciton states in C$_{60}$ are significantly reduced in films containing multiple rotational domains, even when the proportion of non-dominant domains is barely measurable in LEED. When a single domain film is probed by TR-ARPES, the exciton population decay curves obtained are well-modeled with a standard single-exponential decay yielding $\tau_1$ = 8.3 $\pm$ 0.4~ps. An additional, faster decay component must be considered in films that have even a slight hint of multiple domains. This is consistent with previous reports that domain sizes of greater than 20$\times$ the diffusion length of the exciton in a perfectly ordered crystal are necessary to completely mitigate the effects of exciton quenching at grain boundaries in other organic semiconductors\cite{lunt_relationship_2010}. Moreover, a model incorporating two-exciton decay channels is necessary to properly fit the population decay dynamics due to increased local exciton population at grain boundaries. To distinguish between these film qualities (and their associated decay behaviors) a priori, an examination of  highly saturated LEED images for signatures appearing $<$10 dB below the dominant domain peaks is necessary.

We propose that the grain boundaries act as exciton funnels, increasing the local exciton density and therefore increasing two-exciton decay processes such as exciton-exciton annihilation. Measurements of the energy landscape across grain boundaries via STM support this interpretation. In addition, it is likely the reduced structural order at the grain boundary introduces additional, faster radiative and non-radiative decay pathways for excitons via relaxed selection rules and modified phonon coupling. Collectively, these observations are consistent with previous optical studies reporting that increased disorder in organic semiconductors leads to faster quenching of the exciton peaks in photoluminescence\cite{lunt_relationship_2010, sajjad_controlling_2015} and shorter exciton diffusion lengths\cite{terao_correlation_2007, rim_effect_2007}.

These findings highlight the importance of high-quality, well-ordered films for determination of the intrinsic exciton lifetime in organic semiconductors. Moreover, they suggest that disorder-induced exciton quenching may play a critical role in limiting the efficiency of organic optoelectronic devices.

\section{Methods}

The three C$_{60}$ samples (denoted Sample A, B, and C) are grown by MBE on a Au(111) single crystal. The Au(111) substrate is prepared by multiple rounds of Ar$^+$ sputtering followed by annealing at 440$^\circ$C. We can achieve the highest-ordered film by following a two-stage growth method that relies on the difference in the interaction strength between C$_{60}$-Au and C$_{60}$-C$_{60}$ to promote long-range order in the 2$\sqrt{3}\times$2$\sqrt{3}$ R30~$^\circ$ superstructure; details are described elsewhere \cite{tully_two-stage_2024}. The growth rate is calibrated using a water-cooled quartz crystal monitor. Deviations from this optimal growth recipe result in samples with additional rotational domains. For all three samples, LEED is performed at room temperature with a spot size of approximately 250~$\upmu$m. 

Static ARPES measurements are performed using an unpolarized 21.2~eV He I line from a helium discharge lamp with a spot size of 1~mm. TR-ARPES is done with a 3.1~eV pump and a 6.2~eV probe generated from a femtosecond Ti:Sapphire laser system operating at 250 kHz  equipped with a ScientaOmicron DA30-L high-resolution hemispherical electron analyzer\cite{dufresne_versatile_2024}. The probe spot size is 10~$\upmu$m. For all time-resolved measurements, both pump and probe are p-polarized. The pump fluence ranges from 212 to 350~$\upmu$J/cm$^2$ and the time resolution is between 230 and 470~fs for different measurements, as specified. The total energy resolution for each measurement is better than 20~meV. All measurements are performed at 10~K. The MBE, LEED and ARPES chambers are connected, allowing samples to be grown, characterized and measured without exposure to air. The pressure in the entire system is maintained below $1\times 10^{-10}$~mbar.

STM measurements are taken in a separate ScientaOmicron low-temperature scanning probe microscope with a base pressure of $5\times10^{-11}$~mbar, using a Au-terminated etched platinum-iridium tip. All STM measurements are performed at 4.6~K. The sample studied using STM was grown \emph{in situ} in a similar chamber to the MBE, with C$_{60}$ deposited onto a clean Au(111) single crystal at room temperature to intentionally create grain boundaries, while still ensuring the herringbone reconstruction is lifted \cite{altman_nucleation_nodate, gardener_scanning_2009}. 

\begin{acknowledgement}
The authors thank George A. Sawatzky, Ziliang Ye, Scott Oser, Jisun Kim, James Day, and  Hsiang-Hsi Kung for discussions and support.

This research was undertaken thanks in part to funding from the Max Planck-UBC-UTokyo Centre for Quantum Materials, and the Canada First Excellence Research Fund in Quantum Materials and Future Technologies Program. This project is also funded by the Gordon and Betty Moore Foundation's EPiQS Initiative, Grant No. GBMF4779 to A.D. and D.J.J.; the Natural Sciences and Engineering Research Council of Canada (NSERC) Discovery Grants program and QuantaMole Alliance Quantum Consortium; the Canada Foundation for Innovation (CFI); the British Columbia Knowledge Development Fund (BCKDF); the Department of National Defense (DND); the Canada Research Chairs Program (S.A.B. and A.D.); the CIFAR Quantum Materials Program (A.D.); and the NSERC PGSD and Tyler Lewis Clean Energy Research grants program (A.T.).

\end{acknowledgement}

\bibliography{bibliography}

\end{document}